\documentclass[nofootinbib,aps,prl,twocolumn,superscriptaddress,groupedaddress]{revtex4}  
\usepackage{graphicx}  
\usepackage{dcolumn}   
\usepackage{bm}        
\usepackage{amssymb}   

\def\beq{\begin{eqnarray}}
\def\eeq{\end{eqnarray}}

\def\({\left(}
\def\){\right)}

\def\mpl{M_{\rm pl}}

\newcommand{\be}{\begin{equation}}
\newcommand{\ee}{\end{equation}}

\def\ea{\end{eqnarray}}
\def\ba{\begin{eqnarray}}

\def\beq{\begin{eqnarray}}
\def\eeq{\end{eqnarray}}

\def\({\left(}
\def\){\right)}

\def\mpl{M_{\rm Pl}}

\def\lsim{\mathrel{\rlap{\lower3pt\hbox{\hskip0pt$\sim$}}
     \raise1pt\hbox{$<$}}}         
\def\gsim{\mathrel{\rlap{\lower4pt\hbox{\hskip1pt$\sim$}}
     \raise1pt\hbox{$>$}}}         

\def\lsim{\mathrel{\rlap{\lower3pt\hbox{\hskip0pt$\sim$}}
     \raise1pt\hbox{$<$}}}         
\def\gsim{\mathrel{\rlap{\lower4pt\hbox{\hskip1pt$\sim$}}
     \raise1pt\hbox{$>$}}}         

\usepackage{amsmath}
\usepackage{amsfonts}
\usepackage{verbatim}
\usepackage{graphicx}
\usepackage{subfigure}
\usepackage{amssymb}
\usepackage[T1]{fontenc}
\usepackage{slashed}
\usepackage{color}

\begin{document}

\title{Emergent long-range interactions in Bose-Einstein Condensates}
\author{Lasha Berezhiani}
\address{Max-Planck-Institut f\"ur Physik, F\"ohringer Ring 6, 80805 M\"unchen, Germany}
\address{Arnold Sommerfeld Center, Ludwig-Maximilians-Universit\"at, Theresienstra{\ss}e 37, 80333 M\"unchen, Germany}
\author{Justin Khoury}
\address{Center for Particle Cosmology, Department of Physics and Astronomy, University of Pennsylvania,\\ 209 South 33rd St, Philadelphia, PA 19104}
\date{\today}

\begin{abstract}

We consider a massive complex scalar field with contact interactions with a source and show that, upon Bose-Einstein condensation, there is an emergent long-range interaction between sources. This interaction becomes long-range in the limit of vanishing self-interaction between Bose-Einstein constituents. More generally, the range is given by $\ell^{-1}\propto \sqrt{\lambda n/m}$, with $\lambda$ being the 2-body self-interaction coupling constant, $n$ the particle number density in the condensate, and $m$ the mass of the condensed particles. Naively this may sound surprising since in $\lambda\rightarrow 0$ limit gapless excitations of the condensate have dispersion relation $\omega_k=k^2/2m$, yet for the mediated force we have $F\propto 1/r^2$. The reason behind this seemingly counterintuitive result lies in the fact that the force is being mediated by the phonon, which happens to acquire a nontrivial derivative interaction with the source. We discuss the potential ramifications of this observation for dark matter models. In particular, we show that this force  can compete with gravity on galactic scales for a wide range of dark matter mass, provided that the interaction with baryons allows the presence of an extended dark matter condensate core. The effect could be of particular interest in ultra-light dark matter models, such as Fuzzy Dark Matter.

\end{abstract}
\maketitle

{\bf Introduction:} In this Letter, we discuss the emergence of long-range correlations in a Bose-Einstein Condensate (BEC) of massive particles. To the best of our knowledge, Ferrer and Grifols~\cite{Ferrer:2000hm} were the first to find this effect for BEC of non-interacting particles. There, the effective potential between sources (with contact four-point interactions with the BEC building blocks) was obtained using finite-temperature field theory methods. In particular, the interaction between sources was viewed as a loop effect, with the massive degree of freedom (which makes up the condensate) running in the loop. As shown in~\cite{Ferrer:2000hm}, in the presence of a heat bath of ideal gas (with Bose-Einstein statistics) the finite-temperature modification to the propagator of massive particles leads to the emergence of a long-range force. The effect is a bosonic analog of the Kohn-Luttinger effect~\cite{Kohn:1965zz}, which transforms the repulsive contact interaction of fermions into a long-range attractive one in the presence of the Fermi sea. According to~\cite{Kohn:1965zz}, this effect is unrelated to the fermion-phonon attractive interaction in metals, or to van der Waals forces, and is instead due to the sharpness of the Fermi surface. In the absence of sharpness the interaction would be short-range and decay exponentially at large distances.

In this work, we show the emergence of long-range correlations using merely a classical field theory. The advantage of our approach is its simplicity; there is no need to invoke finite-temperature field theory in the analysis. Moreover, the precise nature of statistics is also irrelevant; the analysis relies solely on the presence of a homogeneous bosonic field-configuration. The disadvantage of this method, relative to the one adopted in~\cite{Ferrer:2000hm}, lies in the fact that it can be only applied at temperature well below the critical temperature. However, for cosmological considerations this is precisely the limit we are interested in. On the flip side, our analysis is easily generalizable to self-interacting theories. In fact, we will see that in general the presence of interactions limits the range of the emergent force. Moreover, our approach allows to investigate the case of high-density sources, the presence of which distorts the condensate significantly and as such makes the analysis of~\cite{Ferrer:2000hm} inapplicable. As we will see, because of phenomenological constraints, precisely in the regime of strongly deformed condensate can the long-range interaction between sources compete with gravity.

Besides being an interesting effect in and of itself, the emergent long-range interactions could have important phenomenological consequences for dark matter models in which particles form BECs in galactic halos~\cite{Goodman:2000tg,Hu:2000ke,Hui:2016ltb,Berezhiani:2015pia,Berezhiani:2015bqa,Berezhiani:2017tth}. As one example, we recently proposed a novel dark matter model~\cite{Berezhiani:2015pia,Berezhiani:2015bqa,Berezhiani:2017tth} in which dark matter particles condense into a superfluid phase in the central region of galactic halos. (See also~\cite{Khoury:2016ehj,Hodson:2016rck,Addazi:2018ivg,Cai:2017buj,Fan:2016rda,Alexander:2018fjp,Hossenfelder:2018iym,Sharma:2018ydn,Ferreira:2018wup}.) Phonon excitations mediate a long-range force between baryons, resulting in an effective force law $F\propto 1/r$ that explains various empirical galaxy scaling relations between dark and baryonic components~\cite{Milgrom:1983ca,Famaey:2011kh,Lelli:2017vgz}. This requires coupling baryons to the phase of the dark matter condensate, which explicitly breaks (albeit weakly) the global $U(1)$ symmetry. 

In this paper, instead of mediating a force through the exchange of the phase degree of freedom, let us see what happens if we exchange the modulus of the complex scalar field instead, compatible with the $U(1)$ symmetry. Based on the intuition from the Higgs mechanism, a natural reaction is that this attempt is futile from the get-go. In simple superfluids with bosonic origin there are usually two degrees of freedom in the spectrum: a massless/Goldstone mode and a heavy mode with mass $2m$ corresponding to the particle pair creation. The analogy with the linear sigma model may mislead us to think that the modulus is a heavy mode. However, because the superfluid state breaks Lorentz invariance spontaneously, the situation is more interesting --- both heavy and light modes are some linear combinations of the phase and the modulus of the complex scalar field. As a result, the range of the force mediated by the modulus can be much larger than the Compton wavelength of the heavy mode.

{\bf Theory:} Consider a complex scalar field with quartic self-interaction and $U(1)$-invariant coupling to a source $J$, 
\beq
\mathcal{L}=-|\partial\Phi|^2-m^2|\Phi|^2-\frac{\lambda}{2}|\Phi|^4-\frac{1}{\Lambda^2}|\Phi|^2J\,.
\label{lag}
\eeq
In most of our discussion, we will take the source $J$ to be a density operator, $J = \rho$. It should be stressed that because this is a theory with a mass gap, there are no long-range correlations in vacuum. In particular, if we take two probe sources (introduced as $J$) there will be no long-range $\Phi$-mediated force between them in vacuum. Another obvious reason for the absence of such correlation is the fact that there is only a contact interaction. However, the situation changes dramatically if we submerge these sources within a fluid made of $\Phi$ particles.

It is well known that if $\lambda\geq 0$, $\Phi$ can form a stable homogenous Bose-Einstein condensate. Moreover, if $\lambda$ is strictly positive, then the condensate will exhibit superfluidity,
resulting in the absence of dissipation for subsonic motion. In the semi-classical ({\it i.e.}, mean-field) description, the unperturbed superfluid state is described by a homogeneous classical field configuration, while the excitations about this state are quantized.

Explicitly, in the absence of sources ({\it i.e.}, $J=0$) the superfluid state is described, as usual, by the following solution to the equations of motion
\beq
\Phi=v{\rm e}^{{\rm i}\mu t}\,, \qquad  \mu^2=m^2+\lambda v^2\,.
\label{bg}
\eeq
On this background, the $U(1)$ charge density is given by $n=2\mu v^2$, which corresponds to the particle number density in the superfluid state. In other words, this classical background is determined by one external parameter, $n$.

The spectrum of perturbations around this classical background is also well-known. In particular, because of Goldstone's theorem (see~\cite{Nielsen:1975hm} for the formulation in Lorentz non-invariant theories), we expect one of the two degrees of freedom residing in $\Phi$ to become gapless. A straightforward analysis shows that, indeed, one combination of the modulus and the phase becomes gapless and propagates with dispersion relation
\beq
\omega_k^2\simeq c_s^2 k^2+\frac{k^4}{4m^2}\,,
\eeq
with sound speed $c_s^2\equiv \lambda n/4m^3$. Here we have simplified the (otherwise cumbersome) expression by taking the nonrelativistic limit $k\ll m$ and $c_s^2\ll 1$. The second mode is heavy, with the mass gap $2m$, and as such we will not bother to write the explicit expression. Since we are interested in long-range correlations, this mode will be irrelevant.

{\bf Interaction Between Sources:} As one can see from~\eqref{lag}, the phase of $\Phi$ is decoupled from $J$ due to $U(1)$ invariance. {\it A priori}, based on the intuition from Higgs mechanism, one could naively think that there can be no long-range interactions between sources. However, because of the spontaneous breaking of Lorentz invariance by the background~\eqref{bg}, the modulus of $\Phi$ possesses an admixture with the gapless mode. This makes it possible for the modulus to mediate a force with range much longer than the Compton wavelength~$(2m)^{-1}$ of the heavy mode. 

In order to see how it all works out, let us submerge a point-like source within the condensate. Its presence will obviously distort the homogeneous field configuration~\eqref{bg}. To see this, let us first derive the theory for perturbations. This can be done using the field decomposition
\beq
\Phi=(v+h){\rm e}^{{\rm i}(\mu t+\pi)}\,,
\eeq
where $h$ and $\pi$ represent perturbations of the modulus and phase, respectively. Substituting this back into the action we arrive at
\begin{align}
\label{pertlag}
\mathcal{L}=&~\dot h^2-(\partial_j h)^2+(v+h)^2\Big( \lambda v^2 +2\mu\dot{\pi}+\dot{\pi}^2-(\partial_j\pi)^2\Big)\nonumber\\
&-\frac{\lambda}{2}(v+h)^4-\frac{(v+h)^2}{\Lambda^2}J\,.
\end{align}
At this point, one may wonder about our earlier statement about the heavy mode having a mass gap $2m$, since there is no manifest mass term of this size in the above Lagrangian. The disappearance of the original mass term is due to the fact that the time-dependent phase gives rise to a tachyonic mass, which cancels the original one with $\lambda v^2(v+h)^2$ leftover~\cite{Haber:1981ts}.\footnote{It was shown in~\cite{Dvali:1996zr} that in some theories finite temperature effects could also give a tachyonic contribution to the mass, leading to symmetry non-restoration at high temperature; while here the effect is due to finite density.} However, the main contribution to the gap originates from the kinetic mixing rather than from the mass term.

For simplicity, we are interested in the force mediated by $h$ between point-like, static sources. In the static limit the Lagrangian density reduces to
\begin{align}
\mathcal{L}_{\rm static}=&-(\partial_j h)^2+(v+h)^2\Big( \lambda v^2-(\partial_j \pi)^2 \Big)\nonumber\\
&-\frac{\lambda}{2}(v+h)^4-\frac{(v+h)^2}{\Lambda^2}J\,.
\label{staticlag}
\end{align}
Notice that $\pi$ enters~\eqref{staticlag} quadratically and does not couple to baryons directly. Therefore it can be set to zero without contradicting the classical equations of motion.

For field configurations corresponding to small deviations from the superfluid state, {\it i.e.}, $h\ll v$, we obtain the following linear equation of motion for~$h$
\beq
(\Delta-2\lambda v^2)h=\frac{v}{\Lambda^2}J\,.
\label{h eom linear}
\eeq
It follows that the range of the force is given by
\beq
\ell=\frac{1}{\sqrt{2\lambda v^2}} \simeq \frac{1}{2 m c_s} \gg \frac{1}{2m}\,.
\label{range}
\eeq
The field-configuration sourced by a point-like source, $J=M\delta^{(3)}(\vec{x})$, is
\beq
h(r)=-\frac{v}{\Lambda^2}\frac{M}{4\pi r}{\rm e}^{-r/\ell} \qquad (\text{weak field})\,.
\label{hlinearprofile}
\eeq
Thus for $r\ll \ell$ the force between point-like sources follows an inverse-square law. What is interesting is that the force becomes literally long-range ($\ell\rightarrow\infty$) in the $\lambda\rightarrow 0$ limit. The reason this is somewhat surprising is that we usually associate a $1/r^2$ force with the existence of a massless mediator with dispersion relation $\omega_k\sim k$. In the case of vanishing self-interaction, on the other hand, the gapless dispersion relation is $\omega_k=k^2/2m$, yet the force goes as $1/r^2$. The reason for this lies in the fact that, due to kinetic mixing, the gapless mode enters the amplitude of the complex scalar field with a momentum dependent form factor, as detailed in the Appendix. 

To make this point clear, let us spell out the Lagrangian with $\lambda=0$ at leading order in perturbations 
\beq
\mathcal{L}_{\lambda=0}=\dot h^2-(\partial_j h)^2+\dot{\tilde{\pi}}^2-(\partial_j\tilde{\pi})^2+4mh\dot{\tilde{\pi}}\nonumber\\
-2\frac{v}{\Lambda^2}hJ\,;
\label{linth}
\eeq
where we have redefined the phase by $\tilde{\pi}\equiv\pi v$ for convenience. As we can see, the kinetic part is symmetric in $h$ and $\tilde{\pi}$ up to a mixing term $h\dot{\tilde{\pi}}$. Furthermore, without the latter there would be two gapless modes in the spectrum. In fact, the kinetic mixing is the only term where the mass scale $m$ appears. The observation that $h$ would be a massless mode in the absence of the kinetic mixing, is precisely the reason why it is capable of mediating a long range force between static sources.

Indeed, as detailed in the Appendix, integrating out~$\tilde{\pi}$ and taking the non-relativistic limit results in the effective action
\beq
\mathcal{L}^h_{\lambda=0}=\dot{\hat{h}}^2-\hat{h}\frac{\Delta^2}{4 m^2} \hat{h}-2\frac{v}{\Lambda^2}\left( \frac{\sqrt{-\Delta}}{2m}\hat{h} \right)J\,.
\label{hlinth}
\eeq
where $\hat{h} \equiv \frac{\sqrt{-\Delta}}{2m} h$ is canonically normalized. This expression speaks for itself. The gapless mode (in this formulation described by $h$) mediates the long range force due to nontrivial derivative interaction with the source. This observation indicates that the modulus of the complex field in the original formulation~\eqref{linth} contained the admixture of the gapless mode with a momentum ({\it i.e.}, derivative) dependent form factor.

{\bf Weakly-distorted BEC:} The linear approximation $h\ll v$ underlying the weak-field profile~\eqref{hlinearprofile} is valid on sufficiently large scales, $r \gg r_*$, where
\beq
r_*\equiv\frac{M}{4\pi\Lambda^2}\,.
\eeq
On scales smaller than $r_*$, the superfluid profile is significantly altered by the source, and one must solve the full non-linear equation for $h$. We will do so momentarily. For now, notice that if instead of a point source we considered an object of homogeneous density $\rho$, then its radius $R$ should be greater than $r_*$ in order for the superfluid to be only marginally distorted, and thus for the linear approximation to be valid, everywhere. In other words, such a source should satisfy
\beq
\frac{\rho R^2}{\Lambda^2}<1 \qquad (\text{weak distortion})
\label{rhoRlinear}
\eeq
in order for the condensate to remain in the linear regime.

{\bf Strongly-distorted BEC:} The above discussion immediately raises the question: what happens when sources are dense/large enough to violate~\eqref{rhoRlinear}?  Although an analytic treatment of this problem is in general difficult, fortunately the analysis simplifies in $\lambda\rightarrow 0$ limit. In this limit the condensate's building blocks are non-interacting, which 
incidentally corresponds to the case considered in~\cite{Ferrer:2000hm}. 

For a static source, the equation of motion for $h$ derived from~\eqref{staticlag} reduces to
\beq
\Delta h=\frac{v+h}{\Lambda^2}J\,.
\label{eq}
\eeq
This is a generalization of~\eqref{h eom linear} (albeit with $\lambda = 0$), where $h$ is not assumed small compared to $v$. Assuming a homogeneous density source,
\beq
J=\rho={\rm const.}\qquad \text{for}~r\leq R\,,
\eeq
and $J = 0$ for $r > R$, we can easily find the spherically symmetric classical solution which matches the decaying configuration at infinity and therefore recovers the unperturbed BEC asymptotically. Inside the source, the solution (with non-singular boundary condition at the origin) is
\beq
h_{\rm in}(r)=v \left( -1+\frac{\Lambda {\rm sech}\left[\frac{R\sqrt{\rho}}{\Lambda} \right]}{\sqrt{\rho}}\frac{{\rm sinh}\left[\frac{r\sqrt{\rho}}{\Lambda} \right]}{r} \right)\,.
\eeq
Outside the source the solution is of the long-range form
\beq
h_{\rm out} (r) =-\frac{v}{\Lambda^2}\frac{M_{\rm eff}}{4\pi r}\,,
\eeq
with
\beq
M_{\rm eff}\equiv \frac{4\pi}{3}\rho R^3 \times \frac{3\Lambda^2}{R^2 \rho}\left( 1-\frac{\Lambda}{R\sqrt{\rho}}{\rm tanh}\left[ \frac{R\sqrt{\rho}}{\Lambda} \right]\right)\,.
\eeq
As one can see, the effective mass $M_{\rm eff}$ perceived by a probe located outside the source depends on the density and the size of the source.\footnote{Here, by effective mass we refer to the one sourcing $h$-field. The gravitational field, of course, is unaffected by this screening. In other words, $M_{\rm eff}$ simply represents an ``$h$ hair''.} In particular, it is easy to see that the relevant combination is
\beq
x\equiv \frac{R\sqrt{\rho}}{\Lambda}\,.
\label{x}
\eeq
For low-density sources, $x\ll 1$, the effective mass is very close to the gravitational mass, {\it i.e.},
\beq
M_{\rm eff}\Big|_{x\ll 1}\simeq \frac{4\pi}{3}\rho R^3\qquad (\text{low-density/unscreened}) \,.
\label{lowxmeff}
\eeq
For $x\gg 1$ the effective mass gets significantly screened compared to the gravitational mass, 
\beq
M_{\rm eff}\Big|_{x\gg 1}\simeq 4\pi \Lambda^2 R\qquad (\text{high-density/screened}) \,.
\label{highxmeff}
\eeq
Notice that large-$x$ limit corresponds to small $\Lambda$; in fact, a closer look reveals that $x>1$ corresponds to~\eqref{highxmeff} being smaller than the object's mass.\footnote{This screening effect can be understood intuitively by noticing that, according to~\eqref{eq}, $h$ acquires a mass within the source, which drives symmetry restoration within the source. This is similar to the symmetron screening mechanism~\cite{Hinterbichler:2010es}, albeit in a Lorentz-violating background. In other words, high-density sources basically expel the condensate from their core and recover~$\Phi=0$ background in the center.} In other words, in this high-density limit the most of the source mass is screened, except the thin shell of thickness $\Lambda^2/\rho R$. The higher is the density, the thinner the shell gets. Furthermore, in this limit $h$ becomes independent of the object density (and independent of $\Lambda$):
\beq
h(r)=-v\frac{R}{r} \qquad (\text{high-density/screened}) \,.
\eeq
Thus, remarkably, the $h$-hair of a high-density homogeneous source is its size. This is precisely what happens in the symmetron~\cite{Hinterbichler:2010es} and chameleon~\cite{Khoury:2003aq,Khoury:2003rn} and screening mechanisms, where screening is achieved by making the scalar mediator respectively heavy or weakly-coupled inside dense sources. See~\cite{Joyce:2014kja} for a review of screening mechanisms. Note that the condensate density (equivalently, $v$) merely sets the overall field strength. The field profile and screening effect are completely oblivious to it.

To summarize, there are three interesting limiting cases for the emergent force between two sources submerged within the condensate, depending on their respective densities:

\begin{itemize}

\item [{\bf (i)}] {\bf Unscreened source and unscreened probe:} The force mediated by $h$ between two low-density ($x\ll 1$) sources takes the Newtonian form
\beq
F_h=\frac{v^2}{4 \pi\Lambda^4}\frac{M_1 M_2}{r^2}\,.
\label{F1}
\eeq

\item[{\bf (ii)}] {\bf Screened source and screened probe:} For two high-density ($x\gg 1$) sources, the effective masses can be well approximated by~\eqref{highxmeff}. The resulting force is
\beq
F_h=4\pi v^2\frac{R_1 R_2}{r^2}\,.
\label{F2}
\eeq
Notice that the force is independent of $\Lambda$ in this limit.

\item[{\bf (iii)}]  {\bf Screened source and unscreened probe:} For a system comprised of a high-density source 1, and a low-density source 2, we get
\beq
F_h=\frac{v^2}{\Lambda^2}\frac{R_1M_2}{r^2}\,.
\label{F3}
\eeq
In this case the force still depends on the coupling strength $\Lambda$.
\end{itemize}

The following remark is in order. The above results in the strongly-distorted regime were obtained by setting $J=\rho$ in~\eqref{eq}. In this case, the interaction vertex $\mathcal{L}_{\rm int}=-|\Phi |^2 J/\Lambda^2$ corresponds to a repulsive contact interaction between $\Phi$ and matter quantum in vacuum. On the other hand, if we instead set $J=-\rho$ it would correspond to an attractive interaction. Interestingly, in the weak-field regime $h\ll v$, corresponding to low-density sources, the force is attractive independently of the sign of $\mathcal{L}_{\rm int}$. This is similar to Kohn-Luttinger effect for fermions, and for bosonic system it was pointed out by~\cite{Ferrer:2000hm}. However, in the strongly-distorted regime, corresponding to a homogeneous source with density and size satisfying $R^2\rho/\Lambda^2>1$, a gradient instability emerges in the attractive case, {\it i.e.}, for $J=-\rho$. 

This is easy to see by noticing that inside the source the interaction term generates an additional contribution to the mass term. To see this, let us assume that $\rho/\Lambda^2\ll m^2$, for otherwise the heavy mode residing in $\Phi$ would become unstable. Moreover, even if the heavy mode is stable it is easy to see that the Goldstone boson (gapless mode residing in $\Phi$) acquires the following dispersion relation inside the source
\beq
\label{attrdisp}
\omega_k^2= \pm \frac{\rho}{4\Lambda^2 m^2}k^2+\frac{k^4}{4m^2} \qquad \text{for}~J = \pm \rho\,.
\eeq
Evidently the density-dependent contribution is tachyonic for $J=-\rho$, indicating a gradient instability. Specifically, for $J = -\rho$ perturbations with wavenumber $k<k_*\equiv \frac{\sqrt{\rho}}{\Lambda}$ are unstable.\footnote{For applications to galactic dynamics discussed below, it is easy to argue that the maximal instability rate, achieved for $k \sim k_*$, is much faster than the Hubble rate.} On the other hand,  we can only meaningfully talk about such soft modes as long as the source radius $R$ is larger than $k_*^{-1}$, {\it i.e.}, if $\frac{R\sqrt{\rho}}{\Lambda} = x > 1$. Thus the instability kicks in whenever BEC distortions cease to be linear. 

In other words, in the attractive case, as we crank up the density of the source there is a critical density above which the condensate gets destabilized. In the repulsive case, in contrast, the condensate manages to maintain long-range coherence and exhibits a mass screening discussed above. This resonates with the observed screening effect for repulsive interactions. Indeed, whenever the source repels the condensate degrees of freedom, it is trying to expel the condensate from within itself. Correspondingly $|v+h|$ is always smaller than $v$. In fact, in very dense objects the condensate density practically vanishes at the center of the source. 

This suggests a nice physical interpretation of the screening of $h$-hair. In order to source the hair, the source needs to be submerged within the condensate, as the strength of the $hJ$ coupling depends on the condensate density. Moreover, the condensate is removed (or suppressed) from the location of the source, which in turn results in a weaker $h$-hair. In the attractive case, we have the opposite situation: the source wants to accrete the entire condensate, which results in the aforementioned instability for high-density sources.

{\bf Regime of Validity:} From~\eqref{lag} the interaction between the condensate degrees of freedom and the source is a higher-dimensional operator suppressed by $\Lambda$. Therefore, the theory at hand must be viewed as an effective field theory. Naively the cutoff is $\Lambda$, and to justify neglecting operators of higher dimension one should require 
\beq
v\ll \Lambda\,.
\label{conservative bound}
\eeq
It is interesting to notice that this is the bound we obtained in the weakly-distorted condensate. In the strongly-distorted case, high-density sources suppress the local condensate
density compared to $v$, which makes~\eqref{conservative bound} even easier to satisfy.

Note that demanding $v\ll \Lambda$ is conservative, as the actual strong coupling scale could be much higher than $\Lambda$. For example, suppose that $J$ is the energy-density operator
for a fermion $\psi$ of mass $m_{\rm b}$ (~GeV for a baryon):
\beq
\mathcal{L}\propto \frac{m_{\rm b}}{\Lambda^2}|\Phi|^2\bar{\psi}\psi\,.
\label{element}
\eeq
In this case the actual strong coupling scale of this operator is $\Lambda'=\Lambda^2/m_{\rm b}$. As we will see from phenomenological considerations later, the scale $\Lambda$ will generally be 
a few orders of magnitude below the Planck scale, but much larger than~GeV. Thus we generically expect $m_{\rm b}\ll\Lambda$, which implies $\Lambda' \gg \Lambda$. Nevertheless, when studying the phenomenological implications of our mechanism we will be imposing the conservative bound~\eqref{conservative bound}.

{\bf Dark Matter Context:} The emergent long-range effect discussed above could have interesting ramifications for scalar dark matter models. For concreteness, suppose that $\Phi$ describes dark matter, while $J$ is the trace of the baryonic energy-momentum tensor, $J=-T^\mu_{\;\mu}$. Let us further assume that dark matter forms a Bose-Einstein condensate in the central regions of galaxies, large enough to encompass part (or all) of the baryonic disk (see~\cite{Berezhiani:2017tth}). For simplicity we set $\lambda=0$, to capitalize on the analytical non-linear screening solutions derived earlier.\footnote{It is easy to show that not much is gained by including repulsive self-interactions, $\lambda \neq 0$. The reason is that if self-interactions are responsible for stabilizing the BEC core against gravitational collapse (instead of quantum pressure), then demanding that the range of the emergent force be greater than $\sim 10$~kpc requires unrealistic values for the dark matter particle mass.}
In this case stability is achieved by quantum pressure, which results in a BEC core with radius comparable to the Jeans length:
\begin{align}
L_{\rm J} &=2\pi \left ( \frac{2\rho_{\rm DM}m^2}{\mpl^2} \right)^{-1/4}\nonumber\\
&\simeq 12~{\rm kpc} ~\left(\frac{m}{10^{-22} ~{\rm eV}} \right)^{-1/2} \left( \frac{\rho_{\rm DM}}{10^{-26}~{\rm g/cm}^3} \right)^{-1/4}\,.
\label{jeansscale}
\end{align}
This result, however, neglects the repulsive interaction between baryons and dark matter. More generally, the dispersion relation for density fluctuations around the homogeneous dark matter condensate in the presence of a baryon density $\rho_{\rm b}$ is
\beq
\omega_k^2=-4\pi G_{\rm N} \rho_{\rm DM}+\frac{\rho_{\rm b}}{4\Lambda^2m^2}k^2+\frac{k^4}{4m^2}\,,
\label{findisp}
\eeq
This generalizes~\eqref{attrdisp} for repulsive interactions, by including a tachyonic mass term that comes from the mixing with gravity and leads to
Jeans' instability. For the fiducial low-surface brightness (LSB) galaxy considered below,  {\it i.e.}, $\rho_{\rm b} \simeq 3\times 10^{-26} ~{\rm g/cm}^3$, one
can easily show that the second term dominates over the third one for the relevant scales as long as $\Lambda<10^{-4} \mpl$. Thus the extent of the BEC core can be significantly
greater than~\eqref{jeansscale}.

Baryons within the core experience an additional force (on top of gravity) mediated by $h$. From large-scale structure observations, the dark matter cannot be much lighter than $10^{-22}$~eV. Hence~\eqref{jeansscale} implies that, within the simple model under consideration, the emergent force can extend at most up to $\sim 10$~kpc distance from the galactic center. The force is therefore irrelevant in the flat part of the rotation curves of high-surface brightness (Milky Way-like) galaxies, which extend out to $\sim 50$~kpc. However it can be very much relevant for LSB galaxies whose rotation curves typically extend to at most $10~{\rm kpc}$. This is fortuitous, since LSB galaxies arguably pose a more significant challenge to the standard cold dark matter paradigm.

To quantify the strength of this novel force, consider a fiducial LSB galaxy, IC 2574, with total baryonic mass $M_{\rm LSB}\simeq 10^9\,{\rm M_{\odot}}$~\cite{Lelli:2016zqa}. The density profile of baryons can be crudely modeled as a uniform-density sphere of radius $R_{\rm LSB}\simeq 8~{\rm kpc}$. In reality the baryon distribution is of course neither spherical nor homogeneous, hence the results derived below are for illustrative purposes only. A more thorough analysis with realistic baryon distribution will be considered elsewhere.

Whether the force mediated by dark matter degrees of freedom is in the unscreened or screened regime is set by $\Lambda$:

\begin{itemize}

\item \textit{Linear/unscreened regime:} In order for our fiducial LSB galaxy to weakly-distort the condensate, it should satisfy~\eqref{rhoRlinear}. With $\rho_{\rm LSB} = 3 M_{\rm LSB}/4\pi R_{\rm LSB}^3$, this gives
\beq
\frac{3M_{\rm LSB}}{4\pi R_{\rm LSB}\Lambda^2}\lesssim 1\,.
\eeq
Substituting the fiducial values $M_{\rm LSB}\simeq 10^9 {\rm M_{\odot}}$ and $R_{\rm LSB}\simeq 8~{\rm kpc}$, we obtain 
\beq
\Lambda \gsim 10^{-4}\mpl\,.
\label{lambdabound1}
\eeq
In order to compare the dark force to the baryonic gravity, we use the unscreened force of Case~(i), given by~\eqref{F1}. It must be stressed that we are comparing the forces near the edge of the baryonic distribution, so that~\eqref{F1} can be legitimately used. It is easy to see that $F_h$ will be comparable or stronger than gravity if
\beq
\frac{\rho^{(v)}_{\rm DM}\mpl^2}{m^2\Lambda^4}\gsim 1\,,
\label{cond1}
\eeq
where $\rho^{(v)}_{\rm DM} \equiv 2m^2v^2$ denotes dark matter density in the absence of baryons. Assuming $\rho^{(v)}_{\rm DM}\simeq 10^{-25}~{\rm g/cm}^3$, condition~\eqref{cond1} gives us the following bound on the dark matter particle mass
\beq
m\lesssim 3\times 10^{-23}~{\rm eV}\cdot \left(\frac{\Lambda}{10^{-4}\mpl} \right)^{-2}\,.
\eeq
In light of~\eqref{lambdabound1} this requires a dark matter particle lighter than $\sim 10^{-23}$~eV, which is problematic for large-scale structure observations~\cite{Hui:2016ltb}.

\item \textit{Strongly-distorted/screened regime:} The opposite regime where our LSB galaxy strongly distorts the condensate, $\frac{3M_{\rm LSB}}{4\pi R_{\rm LSB}\Lambda^2} > 1$,
corresponds to
\beq
\Lambda < 10^{-4}\mpl\,.
\label{lambdabound2}
\eeq
In this case, to compare the $h$-mediated force with gravity we consider Case~(iii), with the dark force given by~\eqref{F3}. This assumes an unscreened probe, which is appropriate for the low-density hydrogen used to trace rotation curves. Once again, demanding the dominance of the dark force near the boundary of the baryonic profile for the screened galaxy, we arrive at
\beq
\frac{4 \pi \rho^{(v)}_{\rm DM}R_{\rm LSB}\mpl^2}{m^2\Lambda^2 M_{\rm LSB}}\gsim 1\,.
\eeq
Substituting $\rho^{(v)}_{\rm DM}\simeq 10^{-25}~{\rm g/cm}^3$, as well as our fiducial LSB parameters $M_{\rm LSB}\simeq 10^9 {\rm M_{\odot}}$ and $R_{\rm LSB}\simeq 8~{\rm kpc}$, 
we obtain 
\beq
m\lesssim 2\times 10^{-23}~{\rm eV}\cdot \left( \frac{\Lambda}{10^{-4}\mpl} \right)^{-1}\,.
\label{mbound}
\eeq
Thus by lowering $\Lambda$ the dark matter mass can easily be heavier than $10^{-22}~{\rm eV}$ to comply with large-scale structure constraints. Notice that~\eqref{mbound} can be written equivalently as 
\beq
m\Lambda\lesssim 5~{\rm eV}^2\,.
\eeq

\end{itemize}
To summarize, as long as $\Lambda< 10^{-4}\mpl $, such that LSB galaxies are screened, it is possible for the dark force to be strong enough to compete with the baryonic gravitational force near the edge of the baryonic distribution, while having $m$ large enough to avoid obvious conflicts with structure formation.

Two remarks are warranted here. First, it should be emphasized that the screened regime of interest corresponds to the {\it removal} of significant amount of dark matter from within the baryonic source. As discussed earlier, this results from weak repulsive interactions between baryons and dark matter, which is enhanced due to the high degeneracy of the BEC. This is an interesting observation in and of itself, because some LSB galaxies (like IC~2574 considered here) seem to require such a removal mechanism (which, in the standard context, is attributed to feedback) to improve the rotation curve fit~\cite{Oman:2015xda}.

A second remark is that, although we have been comparing the dark force to the gravitational force due to baryons, it is straightforward to repeat the analysis for the gravitational force due to the dark matter as well. In this case we would obtain an additional factor of $\sqrt{M_{\rm b}/M_{\rm DM}}$ on the right-hand side of~\eqref{mbound}, where $M_{\rm DM}$ is enclosed mass at the relevant distance where the comparison is made. This would only marginally tighten the bound.

Since our scenario involves a long-range force between baryons, one must worry about solar system tests of gravity. Because the predictions of the model on such small scales (compared to galactic scales) are more model-dependent, for the purpose of this paper we will content ourselves with making a few general remarks. As already mentioned, the phenomenology of our model bears close resemblance to the symmetron screening mechanism~\cite{Hinterbichler:2010es}, hence a natural possibility is that symmetron screening ensures compatibility with solar system constraints. It has been argued recently that a symmetron force could affect galactic dynamics on $\sim$kpc scales and impact rotation curve observations while being phenomenologically viable in the solar system~\cite{Burrage:2016yjm,OHare:2018ayv,Burrage:2018zuj}. Alternatively, new effects could kick in on small scales. For instance, additional dark matter self-interactions could result in further short-scale screening, as in the Vainshtein mechanism~\cite{Vainshtein:1972sx,Deffayet:2001uk}. See~\cite{Babichev:2011kq} for an example.  

{\bf Summary and Outlook:} In this paper we discussed the explicit example of the emergence of long-range attractive force in Bose-Einstein condensates from contact interactions (either attractive or repulsive). The effect is reminiscent of the Kohn-Luttinger effect for Fermi liquids, albeit for bosons. 
 
Specifically, in the theory described by~\eqref{lag}, sources in vacuum would only have short-range contact interactions. In the presence of a BEC of $\Phi$-particles, however, there is an emergent attractive interaction between sources mediated by the exchange of $\Phi$. The range of this interaction~\eqref{range} depends on the self-interaction coupling strength and becomes strictly long-range in the absence of self-interaction. This is at face value a curious fact, since precisely in this limit the BEC stops exhibiting superfluidity. Namely, in the limit of negligible interactions, phonons stop propagating as waves with $\omega_k=c_s k$, and instead propagate as gapless particles with $\omega_k=k^2/2m$. The resolution is that, despite having a higher-gradient dispersion relation, phonons also couple derivatively to sources. This is due to kinetic mixing, which results in the modulus of the complex field containing an admixture of the gapless mode with nontrivial momentum-dependent form factor. The end result is an inverse-square law. 

An emergent long-range force in the presence of heat bath was previously pointed out by~\cite{Ferrer:2000hm}. However, their consideration was limited to the non-interacting bosons,
and the effect was established as a thermal loop effect for a Bose-Einstein distribution, obtained by analyzing the asymptotic scattering amplitude. A long-range force was found below critical temperature, though its regime of validity was not analyzed. 

In this work, we have rediscovered the effect using a much simpler method. In the case of BEC of non-self-interacting particles ({\it i.e.}, $\lambda=0$), we have performed a thorough analysis of the emergent long-range correlation and have established the following:

\begin{itemize}
\item For point sources submerged in a BEC, we have found an effective inverse-square force $F\propto 1/r^2$, thus verifying the result of~\cite{Ferrer:2000hm}.

\item Taking into account that the generated force lines in the presence of the source distort the condensate, we have shown that there is a shortest distance down to which the expression for the force may be trusted. For instance, when dealing with the point source the classical field configuration is obtained by matching the singularities at the location of the source, however since the linear approximation breaks down at a finite distance from the source it is unclear to what extent one can trust the matching procedure.

\item Going beyond the analysis of~\cite{Ferrer:2000hm}, we have considered a finite-size spherical source of homogeneous density. This allowed us to solve for the BEC profile in the presence of such a source beyond the weak-distortion regime. We found that the weakly-distorted profile is valid provided that 
\beq
\frac{M}{R\Lambda^2}\ll 1\,;
\label{ineq1}
\eeq
Clearly the limit of a point source $R\rightarrow 0$ keeping $M$ fixed grossly violates this inequality.

\item For sources that violate~\eqref{ineq1} and therefore strongly distort the condensate, we have found that the condensate density is diminished inside the source because of repulsive interactions.\footnote{As we have seen, in the case of attractive interactions between the source and $\Phi$, the condensate is destabilized by sources violating~\eqref{ineq1}.} In fact, for $\frac{M}{R\Lambda^2}\gg 1$ the BEC density in the center of the source practically vanishes. The resulting profile for $h$ in this regime is
\beq
h(r)=-v\frac{R}{r}\,,
\eeq
which, remarkably, is independent of the source mass $M$. This is indicative of screening, more precisely symmetron screening. In the limit of a point source, $R\rightarrow 0$, we see that the distortion in fact vanishes.

\end{itemize}

Furthermore, we have analyzed the impact of self-interactions on the range $\ell$ of the emergent force. In other words, this is the case of a superfluid instead of a BEC. The range at finite quartic coupling $\lambda$ is
\beq
\ell\propto \left(\frac{\lambda n}{m} \right)^{-1/2}\,.
\eeq

As an illustrative application, we considered the implications of our mechanism for BEC dark matter models. In particular, in any theory where dark matter is an ultralight (and weakly self-interacting) scalar particle, a long-range force between baryons should emerge whenever there is s a contact interaction (however weak) between baryons and dark matter. In a broad class of scalar dark matter models, the emergent force can be of comparable strength to gravity provided that parameters satisfy~\eqref{mbound}. 

More precisely, because of screening we have found that the dark force is suppressed deep inside the baryonic distribution, due to the reduction of the dark matter density within the source, 
whereas near the edge of the source the dark force is strong enough to compete with gravity. We should stress that this result was derived for a spherical baryonic profile, which is of
course a crude approximation to a galactic disk geometry. We expect the repulsion effect to be less significant in more realistic disk-like distributions, which will be studied elsewhere. 
Because of this, the phenomenological analysis presented here is for illustrative purposes and should not be taken too literally.

The long-range force discussed in this work could be of particular relevance to explain empirical galactic scaling relations, such as the mass discrepancy acceleration relation.  
In contrast to a fundamental modification of gravity, which would be subject to stringent constraints on various scales, an effective force emerging from dark matter dynamics can be
confined to galactic scales, thereby alleviating possible concerns with universal fifth forces. We should caution that the emergent force is an inverse-square law, $F\propto 1/r^2$, 
whereas galactic scaling relations prefer an ultra long-range $F\propto 1/r$ force. Thus the current work should be viewed as a stepping stone in the search for a suitable emergent force
that can explain galactic scaling relations. We will explore generalizations of our model along these lines elsewhere.

\textit{Comment added ---} While putting the finishing touches to this paper we came across some recent works~\cite{Alby:2017dzl,Hees:2018fpg} on the subject. In particular, the effect of the emergent long range force of~\cite{Ferrer:2000hm} was rediscovered using classical field theory techniques, in the case of a non-self-interacting, massive, real scalar field with nonlinear interaction with a source. There are some similarities with the present work. Specifically, it was shown in~\cite{Alby:2017dzl} that if the scalar field has an oscillating background with a frequency equal to the mass of the scalar field ({\it i.e.}, if the BEC is formed), the mass term cancels out, resulting in the scalar field mediating a long-range force, albeit with an oscillating pre-factor. Furthermore, the approach of~\cite{Alby:2017dzl} was used in~\cite{Hees:2018fpg} to calculate the scalar profile created by the extended distribution of baryons. There it was shown that, depending on the compactness of the source, the scalar density could get significantly screened or amplified, depending on the sign of the coupling between the scalar field and baryons. 

An important difference with the present work is that we instead considered a complex scalar field. This, as we saw, makes the analysis and the generalization to the self-interacting theory much easier. One immediate consequence is the force between static sources is time-independent in our case, whereas the force mediated by a real scalar has a rapid oscillatory pre-factor that can be removed by averaging over time (corresponding to the non-relativistic limit). Furthermore, for attractive interactions between the scalar field and baryons (would be amplified regime) we showed that the static configuration is unstable in the presence of highly-dense sources, a point that was overlooked in earlier works. Moreover, we showed that the presence of self-interactions (which can be easily incorporated with a complex scalar field), leading to the superfluidity of the condensate, limits the range of the emergent force. The inclusion of self-interactions for the real scalar field is possible but more involved as one must deal with an anharmonic oscillator at the background level. Last but not least, we clarified the physical origin of the long-range force, as the consequence of phonon exchange with 
higher-derivative gradient term and derivative coupling to the source.

\section*{Achnowledgements}
We would like to thank Gia Dvali for useful discussions. J.K. is supported in part by the US Department of Energy (HEP) Award DE-SC0013528, NASA ATP grant 80NSSC18K0694, the Charles E. Kaufman Foundation of the Pittsburgh Foundation, and a W.~M.~Keck Foundation Science and Engineering Grant. 

\section*{Appendix}
\renewcommand{\theequation}{A-\Roman{equation}}
\setcounter{equation}{0} 

In this Appendix we further elaborate on the origin of the emergent long range force and discuss its connection with the massless excitation of the condensate. 
For simplicity, we begin with the non-interacting case ($\lambda = 0$), for which the emergent force is truly long range. We will then come back and switch on 
$\lambda$ to show how it alters the argument.

Our starting point is the linearized theory~\eqref{linth} for the condensate degrees of freedom:
\beq
\mathcal{L}_{\lambda=0}=h\Box h+\tilde{\pi} \Box \tilde{\pi}+4mh\dot{\tilde{\pi}} -2\frac{v}{\Lambda^2}hJ\,;
\label{lintha}
\eeq
where, as before, $\tilde{\pi}\equiv\pi v$. The analysis simplifies if we integrate out one of the two degrees of freedom out. One would naturally be inclined to 
integrate out the modulus $h$, since it usually propagates the heavy mode. However, as we will see, the physics is more transparent if one instead solves for $\tilde{\pi}$.

For pedagogical purposes, let us first see what happens if we integrate out $h$. This can be done, at tree-level, using its equation of motion:
\beq
h=\frac{1}{-\Box}\left( 2 m \dot{\tilde{\pi}}-\frac{v}{\Lambda^2} J \right)\,.
\label{hsoln}
\eeq
Substituting this back into~\eqref{lintha}, we arrive at the following Lagrangian density
\beq
\mathcal{L}^{\tilde{\pi}}_{\lambda=0}=\dot{\tilde{\pi}} \frac{4m^2}{-\Box} \dot{\tilde{\pi}}+\tilde{\pi}\Box \tilde{\pi}-4\frac{mv}{\Lambda^2}J\frac{1}{-\Box}\dot{\tilde{\pi}}+\frac{v^2}{\Lambda^4}J\frac{1}{-\Box}J\,.\nonumber \\
\label{higher der 1}
\eeq
Evidently in this formulation the theory is of a higher-derivative nature. This is hardly surprising, as we are trying to describe two Lagrangian degrees of freedom using a single real scalar field $\tilde{\pi}$. Nevertheless, let us press on and bring the theory to a canonical form using the field redefinition
\beq
\tilde{\pi}\rightarrow \frac{\sqrt{-\Box}}{2m}\tilde{\pi}\,.
\label{redef}
\eeq
The result is
\beq
\mathcal{L}^{\tilde{\pi}}_{\lambda=0}=\dot{\tilde{\pi}}^2-\tilde{\pi} \frac{\Box^2}{4m^2} \tilde{\pi}-2\frac{v}{\Lambda^2}\left( \frac{\dot{\tilde{\pi}}}{\sqrt{-\Box}} \right)J+\frac{v^2}{\Lambda^4}J\frac{1}{-\Box}J\,.\nonumber \\
\eeq
Until now we have not made any approximation that would freeze out the heavy mode. As such the condensate sector propagates two dynamical degrees of freedom. Indeed, the propagator of $\tilde{\pi}$ has two poles: one gapless and one with gap $2m$. 

At this point we focus on low energy excitations with $k\ll m$, which implies that the frequency of the gapless mode satisfies $\omega_k\ll k\ll m$. This freezes out the heavy mode, since its on-shell production is energetically forbidden. In this limit the effective theory reduces to
\beq
\mathcal{L}^{\tilde{\pi}}_{\lambda=0}=\dot{\tilde{\pi}}^2-\tilde{\pi} \frac{\Delta^2}{4m^2} \tilde{\pi}-2\frac{v}{\Lambda^2}J\frac{1}{\sqrt{-\Delta}}\dot{\tilde{\pi}}+\frac{v^2}{\Lambda^4}J\frac{1}{-\Delta}J\,,\nonumber \\
\label{append interm}
\eeq
where we have assumed the source is non-relativistic as well. Of particular interest is the last term, which describes a long-range interaction between sources. We stress this was obtained by
integrating out the modulus $h$, which had an admixture of the gapless mode due to kinetic mixing. Naively this long-range effect does not appear to be related to $\tilde{\pi}$, since the latter's coupling to the source is separately present. Furthermore, since the $\tilde{\pi}$-source coupling involves a time derivative, at face value it would seem that  $\tilde{\pi}$ is incapable of mediating any interaction between static sources. 

The correct interpretation is more subtle. In reality, the full long-range effect is the result of contributions from the last two terms in~\eqref{append interm}. Although we have chosen to express the interaction in this way, this does not change the fact that the long-range force is mediated by the gapless (phonon) mode. 

To convince the reader of the validity of this statement, we present an alternative derivation of the low-energy effective action where long-range interactions are manifestly mediated by phonons.\footnote{Needless to say, the two derivations are completely equivalent, up to a source-dependent field redefinition.} For this purpose, let us go back to~\eqref{lintha} and instead integrate out $\tilde{\pi}$. 
It will become clear in a moment that this approach is even simpler. Solving the equation of motion for $\tilde{\pi}$, we obtain
\beq
\tilde{\pi}=\frac{2m\dot{h}}{\Box}\,.
\eeq
This is similar to~\eqref{hsoln}, except that $J$ does not partake in it. Substituting this back into the Lagrangian, we get
\beq
\mathcal{L}^h_{\lambda=0}=\dot{h} \frac{(2m)^2}{-\Box} \dot{h}+h\Box h-2\frac{v}{\Lambda^2}hJ\,.
\eeq
Just like~\eqref{higher der 1}, the higher-derivative nature of this formulation follows from the fact that it describes two Lagrangian degrees of
freedom with one real scalar field. Performing a field redefinition similar to~\eqref{redef}, $\hat{h} = \frac{\sqrt{-\Box}}{2m} h$, and taking the
non-relativistic limit which freezes out the gapped mode, we arrive at
\beq
\mathcal{L}^{\hat h}_{\lambda=0}=\dot{\hat{h}}^2-\hat{h}\frac{\Delta^2}{4 m^2} \hat{h}-2\frac{v}{\Lambda^2} J \frac{\sqrt{-\Delta}}{2m}\hat{h} \,.
\label{hlin}
\eeq
The origin of the long-range force is now clear. In this formulation the only dynamical degree of freedom is a phonon with dispersion relation $\omega_k=k^2/2m$. Furthermore, this mode couples to
the source with a momentum-dependent form factor. The combination of the quadratic dispersion relation and momentum-dependent coupling results in the gapless mode
mediating a $1/r^2$ force. 

We conclude by illustrating the effect of finite $\lambda$ on the above argument. As is well known, self-interactions generate a sound speed for phonons. In fact it is straightforward to show that~\eqref{hlin} becomes
\beq
\mathcal{L}^{\hat h}_{\lambda>0}=\dot{\hat{h}}^2-\hat{h}\left(-c_s^2\Delta+\frac{\Delta^2}{4 m^2} \right)\hat{h}-2\frac{v}{\Lambda^2}J \frac{\sqrt{-\Delta}}{2m}\hat{h}\,,\nonumber \\
\label{hlinlambda}
\eeq
with $c_s^2\equiv \lambda v^2/2m^2$. It follows that the range of interaction shrinks to $\ell=(2mc_s)^{-1}$, despite the fact that the mediator remains gapless.




\end{document}